\documentclass[a4paper]{llncs}

\usepackage[T1]{fontenc}

\usepackage{microtype}
\usepackage{graphicx}
\usepackage{epstopdf}
\usepackage{amsfonts}
\usepackage{url}
\usepackage{multirow}
\usepackage{verbatim}
\usepackage{algorithm}
\usepackage{algorithmic}
\usepackage{listings}
\usepackage{balance}
\usepackage{color}
\usepackage{cuted}
\usepackage{color}
\usepackage{pgf}
\usepackage{tikz}
\usepackage{paralist}
\usetikzlibrary{arrows,shadows,shapes.geometric,snakes}
\usetikzlibrary{automata}

\usepackage{graphicx}
\usepackage{caption}
\usepackage{subcaption}
\usepackage{wrapfig}

\DeclareFontFamily{OT1}{pzc}{}
\DeclareFontShape{OT1}{pzc}{m}{it}{<-> s * [1.10] pzcmi7t}{}
\DeclareMathAlphabet{\mathpzc}{OT1}{pzc}{m}{it}

\usepackage{float}
\floatstyle{ruled}
\newfloat{pseudocode}{t}{psc}
\floatname{pseudocode}{Algorithm}

\definecolor{Brown}{cmyk}{0,0.81,1,0.60}
\definecolor{OliveGreen}{cmyk}{1,1,1,1}
\definecolor{CadetBlue}{cmyk}{0.9,0.9,0.9,0.9}
\definecolor{red}{rgb}{1,0,0}
\definecolor{green}{rgb}{0,1,0}

\newcommand{\descr}[1]{\smallskip \noindent \textbf{#1}}

\usepackage{xspace}

\DeclareCaptionLabelFormat{andtable}{#1~#2  \&  \tablename~\thetable}

\renewcommand\paragraph[1]{{\medskip\noindent \bf #1:}}  

\DeclareCaptionLabelFormat{andtable}{\tablename~\thetable}

\begin{document}


\title{When Privacy meets Security:\\
Leveraging personal information for password cracking
}

\author{M. D\"{u}rmuth\inst{1} \and A. Chaabane\inst{2}\and D. Perito \inst{2}
\and C. Castelluccia\inst{2}}
\institute{Ruhr-University Bochum \\
\email{markus.duermuth@rub.de}
\and
INRIA, France\\
\email{firstname.lastname@inria.fr}
}


  

\maketitle

\thispagestyle{empty}

\begin{abstract}
  Passwords are widely used for user authentication and, despite their weaknesses, will
  likely remain in use in the foreseeable future. Human-generated passwords typically
  have a rich structure, which makes them susceptible to guessing
  attacks. In this paper, we study the effectiveness of guessing attacks
  based on Markov models. 
  Our contributions are two-fold.
  First, we propose a novel password cracker based on Markov models, which
  builds upon and extends ideas used by Narayanan and Shmatikov (CCS 2005). In
  extensive experiments we show that it can crack up to 69\% of
  passwords at 10 billion guesses, more than all probabilistic password crackers
  we compared against. Second, we systematically analyze the idea that 
  additional personal information about a user helps in speeding up
  password guessing.  We find that, on average and by carefully
  choosing parameters, we can guess up to 5\% more passwords, especially
  when the number of attempts is low. Furthermore, we show that the gain can go up to 30\% for 
  passwords that are actually based on personal attributes. These passwords are clearly weaker and should
  be avoided. Our cracker could be used by an organization to detect and reject them.
  To the best of our knowledge, we are the first to systematically
  study the relationship between chosen passwords and users' personal
  information.  We test and validate our results over a wide
  collection of leaked password databases.
\end{abstract}




\pagestyle{plain}

\section{Introduction}
\label{sec:introduction}


Password-based authentication is the most widely used form of user
authentication, both online and offline.  Passwords  will likely remain the
predominant form of authentication for the foreseeable future, due to  a number
of advantages: passwords are highly portable, easy to understand for
laypersons, and easy to implement for the operators.
Despite the weaknesses passwords have, they still are and will be in
use for some time.  The reason can be found
in~\cite{bonneau-12-replace-passwords}, which lists a large number of
criteria that user authentication may fulfill, and measures the
quality of a large number of user authentication mechanisms.
Alternative forms of
authentication can complement password, but have not been
able, so far, to provide a standalone alternative solution. 


In this work, we concentrate on offline guessing attacks, in which the attacker can 
make a number of guesses bounded only by the time and resources she is willing to invest.
While such attacks can be improved by increasing the speed with which
an attacker can make guesses (e.g., by using specialized hardware and 
large computing resources~\cite{Kedem:1999:BFA:1251421.1251429,hashcat}), we concentrate here on techniques
to reduce the number of guesses required to crack a password. Hence, our approach reduces the attack time independently of the available resources.
 
The optimal strategy for password cracking (both offline and online) is to enumerate passwords in decreasing order of likelihood, i.e., trying more frequent passwords first and less frequent passwords later. Moreover, human chosen passwords frequently have a rich structure which can be exploited to generate candidate guesses (e.g., using a dictionary and a set of concatenation rules).


Tools commonly used for password cracking, such as John the Ripper (JtR),
exploit regularities in the structure of password by applying \emph{mangling rules}
to an existing dictionary of words (e.g., by replacing the letter \texttt{a} with \texttt{@}  or by appending a number). Such approach, allows to generate new guesses from an existing corpus
of data, like a dictionary or a previously leaked password database.
%
Recent work~\cite{weir-2009-pwd-crack-grammars}[13] has shown ways to improve cracking
performance by enumerating guessed passwords based on their likelihood.
These password crackers have been shown to outperform JtR in certain conditions.
The first insight of our work will be to build upon and improve on the
performance of these probabilistic password crackers. 
Furthermore, while previously proposed password crackers outperform JtR, they do not consider any {\em user specific}
information. This means that the guesses outputted by each of these tools
are fixed and do not depend upon additional information about the user.\footnote{JtR does some very limited guessing depending on the username}

Common sense would suggest that guessing a password can be done more
efficiently when personal information about the victim is
known. For example, one could try to guess passwords that contain the victim's date of birth
or the names of their siblings. 
However, this raises the question, how do we order (in a probabilistic sense) what personal information to 
include in the guessing? While guessing likely passwords, shall we include all the information
about the victims or only restrict the attack to a subset of that information? 
To the best of our knowledge, these questions have not been throughly explored so far and, as we will show, have a serious impact on the security of passwords.  
This is especially true since the steadily increasing use of social networks
gives attackers access to a vast amount of public information about their victims for
the purpose of password cracking.

\paragraph{Paper organization}
We review some basics on password guessing, commonly used password
guessers, as well as more related work in
Section~\ref{sec:password-security}.
In Section~\ref{sec:markov-cracking-algos} we describe the
\emph{Ordered Markov ENumerator} (OMEN), and compare its performance
with other password guessers.
Section~\ref{sec:personal-info} details the idea of
exploiting personal information in password guessing, some basic
statistics about the data we use, 
a detailed description of our algorithm and the results of our experiments.
%
We finally conclude this paper with a discussion of our findings.

\section{Related Work}
\label{sec:password-security}

%
%
%

One of the main problems with passwords is that many users choose {\em weak} passwords.  
These passwords typically
have a rich structure and thus can be guessed much faster than with brute-force guessing attacks.
Best practice mandates that only the hash of a password is stored on
the server, not the password, in order to prevent
leaking plain-text when the database is compromised.  Furthermore,
additional \emph{salting} is used to avoid pre-computation attacks.
Let $H$ be a hash function and $\mathit{pwd}$ the password, choose a
random bitstring $s \in^\mathcal{R} \{0,1\}^{16}$ as salt and store
the tuple $(s, h = H(\mathit{pwd~||~s})).$
In this work we mainly consider \emph{offline guessing attacks}, where
an attacker is given access to the tuple $(s,h)$, and tries to recover
the password $\mathit{pwd}$.  
The hash function is frequently designed for the purpose of slowing down
guessing attempts~\cite{Provos:1999:FPS:1268708.1268740}. 
This means that the cracking effort is \emph{strongly dominated by the computation
of the hash function} making the cost of generating a new guess relatively small.
Therefore, we evaluate all password crackers based on the number of attempts they make
to correctly guess passwords.

\subsection{John the Ripper}

John the Ripper (JtR)~\cite{JtR} is one of the most popular password crackers. It proposes
different methods to generate passwords:
In {\em dictionary} mode, a dictionary of words is provided as input, and the tool tests each one of them. Users can also specify various
mangling rules.
When mangling rules are provided, JtR applies each rule to each
word in the input dictionary. 
In real attacks, the dictionary mode using simple mangling rules works surprisingly well, especially when the input
dictionary is derived from large collections of leaked passwords.
Similarly to~\cite{Dell'Amico:2010:PSE:1833515.1833671}, we discover that for relatively small number of guesses (less than $10^8$),
JtR in dictionary mode produces best results.
However, we focus on attacks with larger number of attempts, for which simple, non probabilistic approaches fall short.
%



\subsection{Password Guessing with Markov Models}
\label{sec:markov}

Markov models have proven very useful for computer security in general
and for password security in particular.  
They are an effective tool to crack passwords~\cite{ccs05}, and can
likewise be used to accurately estimate the strength of new
passwords~\cite{castellucia-12-pwd-strength}.
%


The underlying idea is that adjacent letters in human-generated
passwords are not independently chosen, but follow certain
regularities (e.g., the $2$-gram \texttt{th} is much more likely than
\texttt{tq} and the letter \texttt{e} is very likely to follow
\texttt{th}).
In an $n$-gram Markov model, one models the probability of the next
character in a string based on a prefix of length $n-1$.
%
Hence, for a given string $c_1,\ldots,c_m$, a Markov model estimates
its probability as
\begin{equation}
  \label{eq:markov}
  P(c_1,\ldots,c_m)\\ 
  \quad = P(c_1,\ldots,c_{n-1}) \cdot
  \prod_{i=n}^m  P(c_i|c_{i-n+1},\ldots,c_{i-1}). \nonumber
\end{equation}
%
%
%
For password cracking, one basically learns the initial probabilities
$P(c_1,\ldots,c_{n-1})$ and the transition probabilities
$P(c_n|c_{1},\ldots,c_{n-1})$ from real-world data (which should be as
close as possible to the distribution we expect in the data that we
attack), and then enumerates passwords in order of descending
probabilities as estimated by the Markov model (according to
Equation~\ref{eq:markov}). 

%
To make this attack efficient, we need
to consider a number of details. First, one usually has a limited dataset when learning the
initial probabilities and transition probabilities.  The limited data
entails that one cannot learn frequencies with arbitrarily high
accuracy, i.e., \emph{data sparseness} is a problem.  The critical
parameters are the size of the alphabet $\Sigma$ and the parameter
$n$, which determines the length of the $n$-grams.

Second, one needs an algorithm that \emph{enumerates the passwords} in
the right order.  While it is easy to compute the probability for a
given password, it is not clear how to enumerate the passwords in
decreasing probability. To overcome this problem, 
~\cite{ccs05} provides a method to enumerate all
passwords that have probability larger than a given threshold
$\lambda$, but not necessarily in descending order.
Hence, {\em all} passwords that have a probability
(approximately) higher than $\lambda$ are produced in output, where
$\lambda$ is an input parameter. This is sufficient for attacks based on precomputation (rainbow tables), but not for ``normal'' guessing attacks where guessing passwords in the correct order can drastically reduce
guessing time.

Notably, an add-on to JtR has been released with an independent implementation
of the algorithm presented in~\cite{ccs05}. The implementation is available
at~\cite{JtR}. In the evaluation section, we use this implementation as a comparison (and refer to as JtR-markov).

\paragraph{JtR Incremental mode (JtR-inc)~\cite{JtR}} The {\em incremental} mode tries passwords based on a (modified) $3$-gram Markov model. Specifically, JtR-inc computes not only the probability of each $3$-gram but also the probability that this particular $3$-grams appears at certain indices. In this way, this attack takes into account the structure of the password (e.g., upper case appears usually at the front while numbers at the end). However, two key differences are to note: first, as for Narayanan and al. algorithm, the guesses are not generated in the \emph{true} probability order and second, the Markov chain is modified so that it can cover the entire keyspace.

%
%

\subsection{Probabilistic Grammars-based Schemes}


%

Probabilistic context-free grammar (PCFG) schemes make the assumption that password structures have different 
probabilities~\cite{weir-2009-pwd-crack-grammars}. In other words, some structures, such as
passwords composed of 6 letters followed by 2 digits, are more frequent than others.
The main idea of these schemes is therefore to  extract the most frequent structures, and
use them to generate password candidates.

More precisely, in the \emph{training phase}, different structures are
extracted from lists of real-world passwords, where each structure
indicates the positions of lower and uppercase letters, numerical, and
special characters, as well as the associated probabilities. In the \emph{attack phase (or password generation phase)}, an algorithm outputs the
possible structures for the grammar with decreasing probabilities.
From this output, which describes positions of the four classes of
characters, password guesses are generated as follows: 
Numerical and special characters are substituted by those that have
been observed in the training phase and in decreasing order of
probability, and letters are substituted with appropriate words from a
dictionary. This gives the final password candidate list.

\section{OMEN: An Improved Markov Model Based Password Cracker}
\label{sec:markov-cracking-algos}

In this section we present a more efficient implementation of password
enumeration based on Markov models, which is the first 
contribution of our work. Our implementation improves previous
work based on Markov models by Narayanan et al.~\cite{ccs05} and JtR~\cite{JtR}.
Note that the indexing algorithm presented by Narayanan et al.~\cite{ccs05}
combines two ideas: first, it uses Markov models to index only
 passwords that have high probability,
and second, it utilizes a
hand-crafted finite automata to accept only passwords of a specific
form (e.g., eight letters followed by a digit). 
Our algorithm is solely based on Markov models to create guesses. However,  it could
be combined with similar ideas as well.

\subsection{An Improved Enumeration Algorithm}
\label{sec:markov-improved enumeration}

Narayanan et al.'s indexing algorithm~\cite{ccs05} has the
disadvantage of not outputting passwords in order of decreasing
probability,  however, guessing passwords in the right order can
substantially speed up password guessing (see the example in
Section~\ref{sec:omen-experiments}).
We developed an algorithm, the \emph{Ordered Markov ENumerator}
(OMEN), to enumerate passwords with (approximately) decreasing
probabilities. 

On a high level, our algorithm discretizes all probabilities into a
number of bins, and iterates over all those bins in order of
decreasing likelihood.  For each bin, it finds all passwords that
match the probability associated with this bin and outputs them.
More precisely,                 
we first take the logarithm of all $n$-gram probabilities,
and discretize them into levels (denoted $\eta$) similarly to Narayanan et al.~\cite{ccs05}, according to the
formula
$
\mathit{lvl}_i = \mathrm{round} \left( \log( c_1\cdot \mathit{prob}_i
  + c_2 ) \right),
$
where $c_1$ and $c_2$ are chosen such  that the most frequent $n$-grams
get a level of $0$ and that $n$-grams that did not appear in the
training are still assigned a small probability.  
Note that levels are negative, and we adjusted the parameters to get
10 different levels, i.e., the levels  can take values
$0,-1,\ldots,-9$. The number of levels influences both the accuracy of the algorithm as
well as the runtime: more levels means better accuracy, but increased
runtime.


For a specific length $\ell$ and level $\eta$, $\mathrm{enumPwd}(\eta,\ell)$
proceeds as follows:
\begin{enumerate}
\item It identifies all vectors $\vec a = (a_2,\ldots,a_l)$ of length
  $\ell-1$ (when using $3$-grams we need $\ell-2$ transition probabilities
  and $1$ initial probability to determine the probability for a
  string of length $\ell$), such that each entry $a_i$ is an integer in the range $[0,-9 ]$, 
 and the sum of all elements is $\eta$.

\item For each such vector $\vec a$, it selects all $2$-grams $x_1x_2$
  whose probabilities match level $a_2$.
  For each of these $2$-grams, it iterates over all  $x_3$
  such that the $3$-gram $x_1 x_2 x_3$ has level $a_3$.
  Next, for each of these $3$-grams, it iterates over all $x_4$ such
  that the $3$-gram $x_2 x_3 x_4$ has level $a_4$, and so on, until
  the desired length is reached.
  In the end, this process outputs a set of candidate passwords  of
  length $\ell$ and level (or ``strength'') $\eta$.

\end{enumerate}

  A more formal description is presented in
  Algorithm~\ref{code:enum_pwd}. It describes the algorithm for $\ell=4$. However, 
  the extension to larger $\ell$ is straightforward.

\begin{pseudocode}
  \textbf{function} $\textrm{enumPwd}(\eta, \ell)$
  \begin{compactenum}
  \item for each vector $(a_i)_{2\leq i\leq \ell}$ with $\sum_i a_i = \eta$\\
    and for each $x_1 x_2 \in \Sigma^2$ with $L( x_1 x_2 )=a_2$\\
    and for each $x_3\in \Sigma$  with $L(x_3 ~|~ x_1 x_2 )=a_3$\\
    and for each $x_4\in \Sigma$ with $L(x_4 ~|~ x_2 x_3
    )=a_4$:
    \begin{compactenum}
    \item output $x_1 x_2 x_3 x_4$
    \end{compactenum}
  \end{compactenum}
  \caption{Enumerating passwords for level~$\eta$ and length~$\ell$ (here
    for $\ell=4$).\protect\footnotemark}
  \label{code:enum_pwd}
\end{pseudocode}
\footnotetext{Here $L(x y)$ and $L(z|xy)$ stand for the level of 
  initial and transition probabilities, respectively.}


\paragraph{Example}
We illustrate the algorithm with a brief example.  For simplicity, we
consider passwords of length $\ell=3$ over a small alphabet
$\Sigma=\{a,b\}$, where the initial probabilities have levels
\[
  \begin{array}{ll}
    L(aa) = 0,  
   &L(ab) = -1, \\
    L(ba) = -1, 
   &L(bb) = 0,
\end{array}
\]
and transitions have levels
\[
  \begin{array}{ll}
    L(a|aa) = -1
    & L(b|aa) = -1 \\
    L(a|ab) = 0 
    & L(b|ab) = -2 \\
    L(a|ba) = -1
    & L(b|ba) = -1 \\
    L(a|bb) = 0 
    & L(b|bb) = -2. 
  \end{array}
\]

\begin{itemize}
\item Starting with level $\eta=0$ gives the vector $(0,0)$, which
  matches to the password \texttt{bba} only (the prefix ``aa'' matches
  the level $0$, but there is no matching transition with level $0$).

\item Level $\eta=-1$ gives the vector $(-1,0)$, which yields
  \texttt{aba} (the prefix ``ba'' has no matching transition for level
  $0$), as well as the vector $(0,-1)$, which yields \texttt{aaa} and
  \texttt{aab}.

\item Level $\eta=-2$ gives three vectors: $(-2,0)$ yields no
  output (because no initial probability matches the level $-2$),
  $(-1,-1)$ yields \texttt{baa} and \texttt{bab}, and $(0,-2)$ yields
  \texttt{bba}.
\item and so one for all remaining levels. 
\end{itemize}

The selection of $\ell$ (i.e. the length of the password to be guessed) is challenging, 
as the frequency with which a password length appears in the training data is not a good indicator
of how often a specific length should be guessed. For example, assume that are as many passwords
of  length $7$ and of length $8$, then  the success probability of passwords of 
length $7$ is larger as the search-space is smaller.  Hence, passwords of length $7$ should be guessed first.
Therefore, we  use an adaptive algorithm that keeps track 
of the success ratio of each length and
schedules more passwords to guess for those
lengths that were more effective.
More precisely, our adaptive password scheduling algorithm works as follows:

\begin{enumerate}
\item For all $n$ length values of $\ell$ (we
  consider lengths from $3$ to $20$, i.e. $n=17$),  execute $\mathrm{enumPwd}(0 ,\ell)$ 
  and compute the success probability $\mathit{sp}_{\ell,0}$. This probability is computed as 
the ratio of successfully  guessed passwords over the  number of generated password guesses of length $\ell$.

\item Build a list $L$ of size $n$, ordered by the success
  probabilities, where each element is a triple $(\mathit{sp},
  \mathit{level}, \mathit{length})$.  (The first element $L[0]$
  denotes the element with the largest success probability.)

\item \label{item:loop}
Select the length with the highest success probability, i.e.,
  the first element $L[0]=(sp_0,level_0, length_0)$ and remove it from
  the list.

\item Run $\mathrm{enumPwd}(\mathit{level}_0 - 1, \mathit{length}_0)$,
  compute the new success probability $\mathit{sp}^*$, and add the
  new element $(\mathit{sp}^*, \mathit{level}_0-1, \mathit{length}_0)$
  to $L$.

\item Sort $L$ and go to Step~\ref{item:loop} until $L$ is empty or enough guesses have been
  made.

\end{enumerate}


\subsection{Performance Evaluation}
\label{sec:omen-experiments}
In this section, we present a comparison between our improved Markov model 
password cracker and previous state-of-the-art solutions. 

\paragraph{Datasets}
We evaluate the performance of our password guesser on multiple 
datasets.
One of the largest lists currently publicly available is the
\emph{RockYou list} (RY), consisting of $32.6$ million passwords that
were obtained by an SQL injection attack in 2009.  The passwords were
leaked in clear, all further information was stripped from the list
before it was leaked to the public.
This list has two advantages: first, its large size gives well-trained
Markov models;
second, it was collected via an SQL injection attack therefore affecting all the users of the compromised service.
We split the RockYou list into two subsets: a \emph{training set}
(RY-t) of $30$ million and a \emph{testing set} (RY-e) of the
remaining $2.6$ million passwords.

The \emph{MySpace list} (MS) contains about $50\,000$ passwords
(different versions with different sizes exist, most likely caused by
different sanitation or leaked from the servers at different points in
time).  The passwords were obtained in 2006 by a phishing attack.
%
As before, we split the list in a \emph{training set}
(MS-t) of $30\,000$ and a \emph{testing set} (MS-e) of the
remaining $20\,000$ passwords.

The \emph{Facebook} list (FB)  was posted on the \url{pastebin.com} website\footnote{\url{http://pastebin.com/}} in
2011. This dataset contains both Facebook passwords and associated email
addresses. It is unknown how the data was obtained by the hacker, but most
probably was collected via a phishing attack. Finally, we used a list of
$60\,000$ email addresses and passwords leaked by the group LulzSec (we call
this list LZ).  The list was publicly released in June 2011 via Twitter\footnote{\url{https://twitter.com/#!/LulzSec/status/81327464156119040}}.

We complemented the Facebook list by collecting the public information
associated to the Facebook profiles connected to the email addresses.  
For each profile, we collected the public attributes, which include: first/last name;
location; date of birth; friends names; siblings names; education/work names.

\paragraph{Ethical Considerations}
Leaked password databases have been used in a number of studies on passwords~\cite{weir-2009-pwd-crack-grammars,weir-2010-testing-pwd-metrics,castellucia-12-pwd-strength}.  
Studying databases of leaked password has arguably helped the understanding of
users real world password practices. The information we used in our study was already available to the public.


%
%
%
%
%

\paragraph{Results}
In this section, we evaluate the efficiency of our password guesser OMEN, and compare
it with other password guessers on different datasets. 
We discover that OMEN has consistently better performance compared to previously proposed algorithms.
For the experiments, we trained OMEN using the RockYou training set
RY-t (or MS-t in one experiment), and evaluated it on the sets RY-e,
MS (or MS-e when training on MS-t), and FB.
Table \ref{table:summary} provides a summary of the results.
The table can also be used as a comparison of all the previously proposed
password crackers.
\begin{table}[h!]
\begin{center}
\begin{tabular}{|c|c|c|c|c|}
\hline \multirow{2}{*}{Algorithm} & \multirow{2}{*}{Training Set} & \multicolumn{3}{|c|}{Testing Set} \\ \cline{3-5} 
                                    &  & RY-e & MS-e & FB \\ 
\hline \multirow{3}{*}{Omen}    & RY-t ($10^{10}$)& 69\% & 66\% & 64\% \\ \cline{2-5}
                                & RY-t  & 60\% & 54\% & 54\% \\ \cline{2-5}
                                & MS-t ($0.8*10^{10}$)& 64\% & 68\% & 49\% \\ 
\hline PCFG~\cite{weir-2009-pwd-crack-grammars} &  MS-t ($0.8*10^{10}$)& 37\% & 53\% & 29\% \\ \cline{2-5}
\hline \multirow{2}{*}{JtR-Markov~\cite{ccs05}} & RY-t ($10^{10}$) & 64\% & 60\% &61\% \\ \cline{2-5}
                                & RY-t & 53\% & 40\% &50\% \\ \cline{2-5}
\hline JtR-Inc                  & RY-t & 44\% & 37\% &40\% \\ \cline{2-5}
\hline 
\end{tabular} 
\caption{Summary table indicating the percentage of cracked passwords for 1 billion guesses (or 10 billion when specified).}
\label{table:summary}
\end{center}
\end{table}
\vspace*{-1.3cm}
%
%

%

\begin{figure}[h]
        \centering
        \begin{subfigure}{0.5\textwidth}
                \centering
                \includegraphics[width=\textwidth]{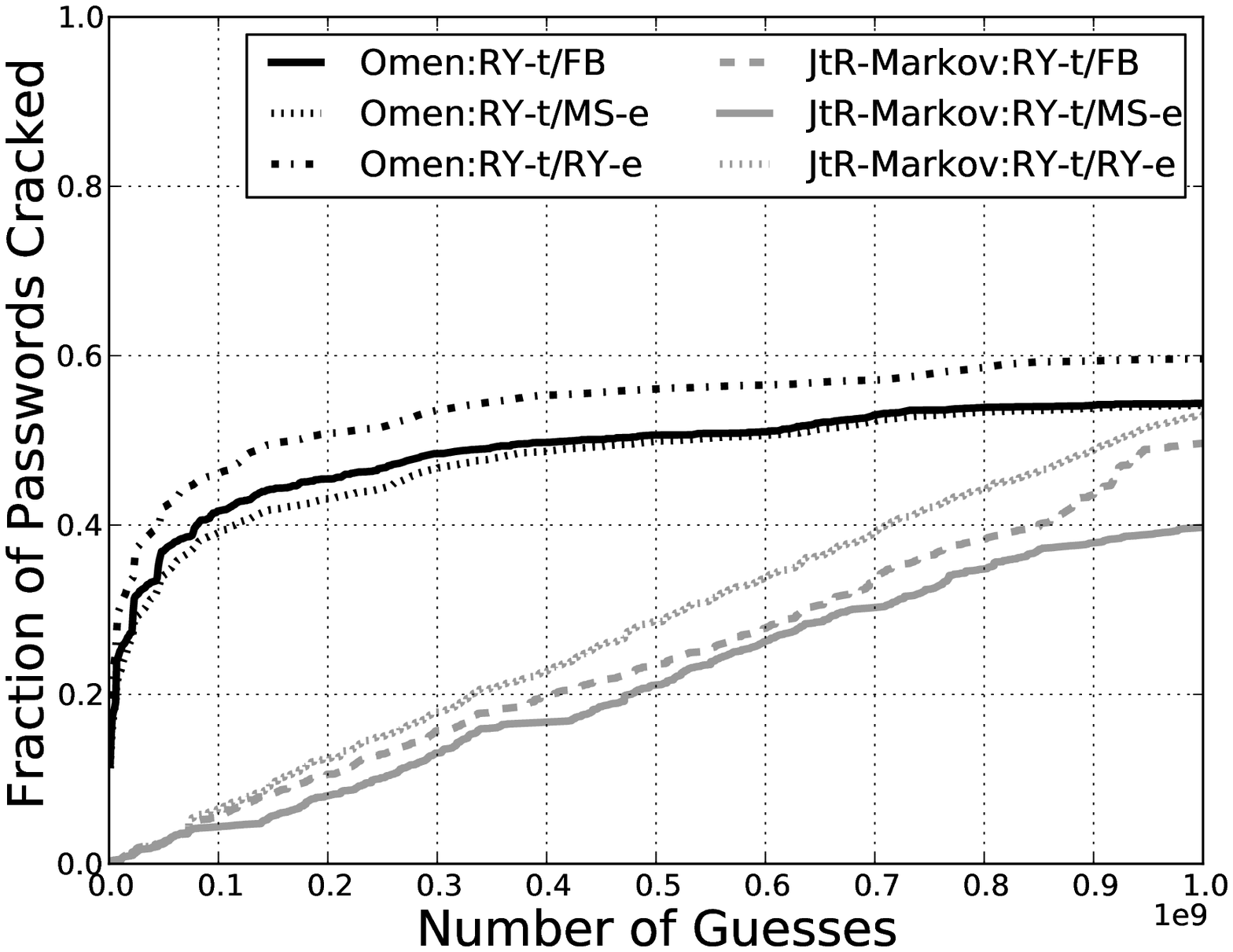}
                \caption{}
                \label{fig:ex-omen-vs-jtr-markov}
        \end{subfigure}%
        ~ 
        \begin{subfigure}{0.5\textwidth}
                \centering
                 \includegraphics[width=\textwidth]{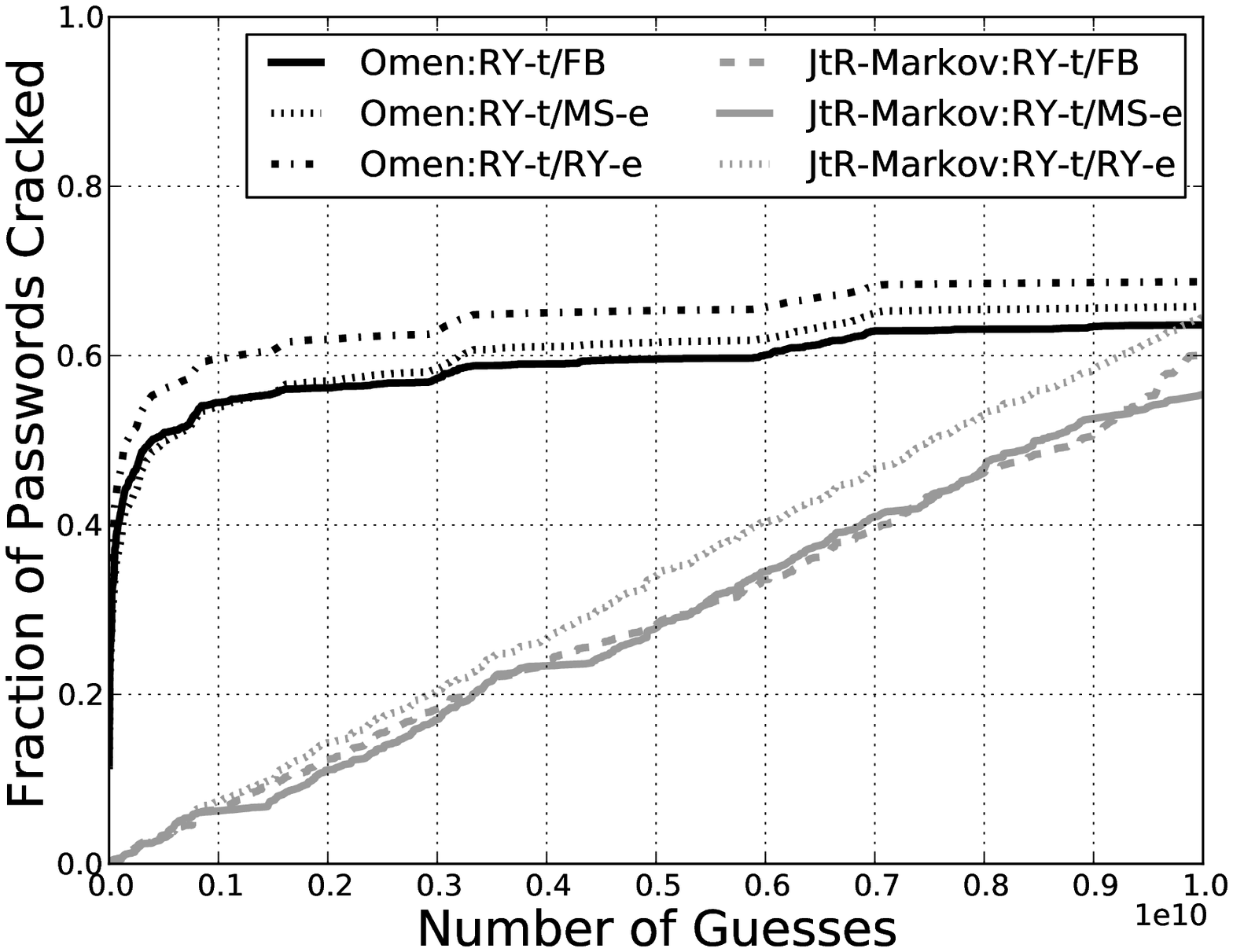}
  				 \caption{}
				  \label{fig:10B-omen-vs-jtr-markov}
        \end{subfigure}
        \caption{(a) Comparing OMEN with the JtR Markov mode, 1B guesses. (b) OMEN Vs JtR Markov mode, 10 billion guesses}
\vspace*{-0.5cm}

\end{figure}

\paragraph{OMEN vs JtR's Markov Mode}
Figure~\ref{fig:ex-omen-vs-jtr-markov} shows the comparison of OMEN
and the Markov mode of JtR.  JtR's Markov mode implements the password indexing function by Narayanan et
al.~\cite{ccs05}.
Both models are trained on a list of passwords (*-t).  Then, given a target
number of guesses $T$ (here 1 billion), we computed the corresponding level ($\eta$) 
to output $T$ passwords. 
The curve shows the dramatic improvement in cracking $speed$ given by our improved
ordering of the password guesses. In fact, JtR-Markov outputs guesses in no particular order which implies
that likely passwords can appear ``randomly'' late in the guesses. This behaviour  leads to the near-linear
curves shown in Figure~\ref{fig:ex-omen-vs-jtr-markov}. One may ask whether JtR-Markov would surpass
OMEN after the point $T$; the answer is \emph{no} as the results do not extend
linearly beyond the point $T$; and larger values of $T$ lead to a flatter curve. To demonstrate this claim, 
we performed the same experiment with $T$ equals to 10 billion guesses (instead of 1 billion). Figure~\ref{fig:10B-omen-vs-jtr-markov}
shows how the linear curve becomes {\em flatter}.

To show the generality of our approach, we compare the cracking performance on three different datasets:RY, FB and MS. 
The ordering advantage allows OMEN to crack more than 40\% of passwords (independently of the dataset) in the first
90m guesses while JtR-Markov cracker needs at least eight times as many guesses to reach the same goal.


%


\begin{figure}[ht!]
\vspace{-0.4cm}
        \centering
        \begin{subfigure}{0.5\textwidth}
                \centering
                 \includegraphics[width=\linewidth]{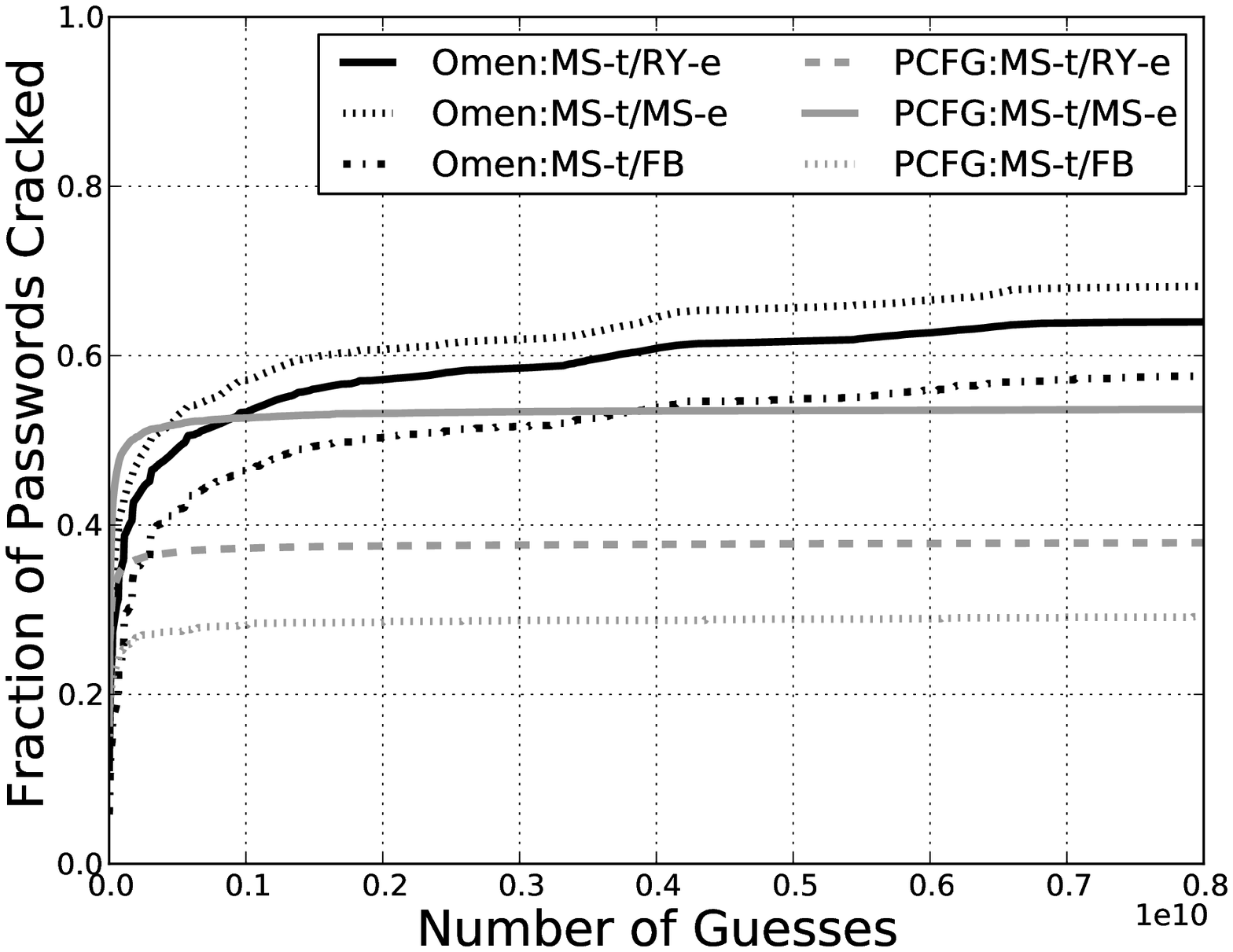}
  				\caption{}
  				\label{fig:ex-omen-vs-pcfg}
        \end{subfigure}%
        ~ 
        \begin{subfigure}{0.5\textwidth}
                \centering
                 \includegraphics[width=\linewidth]{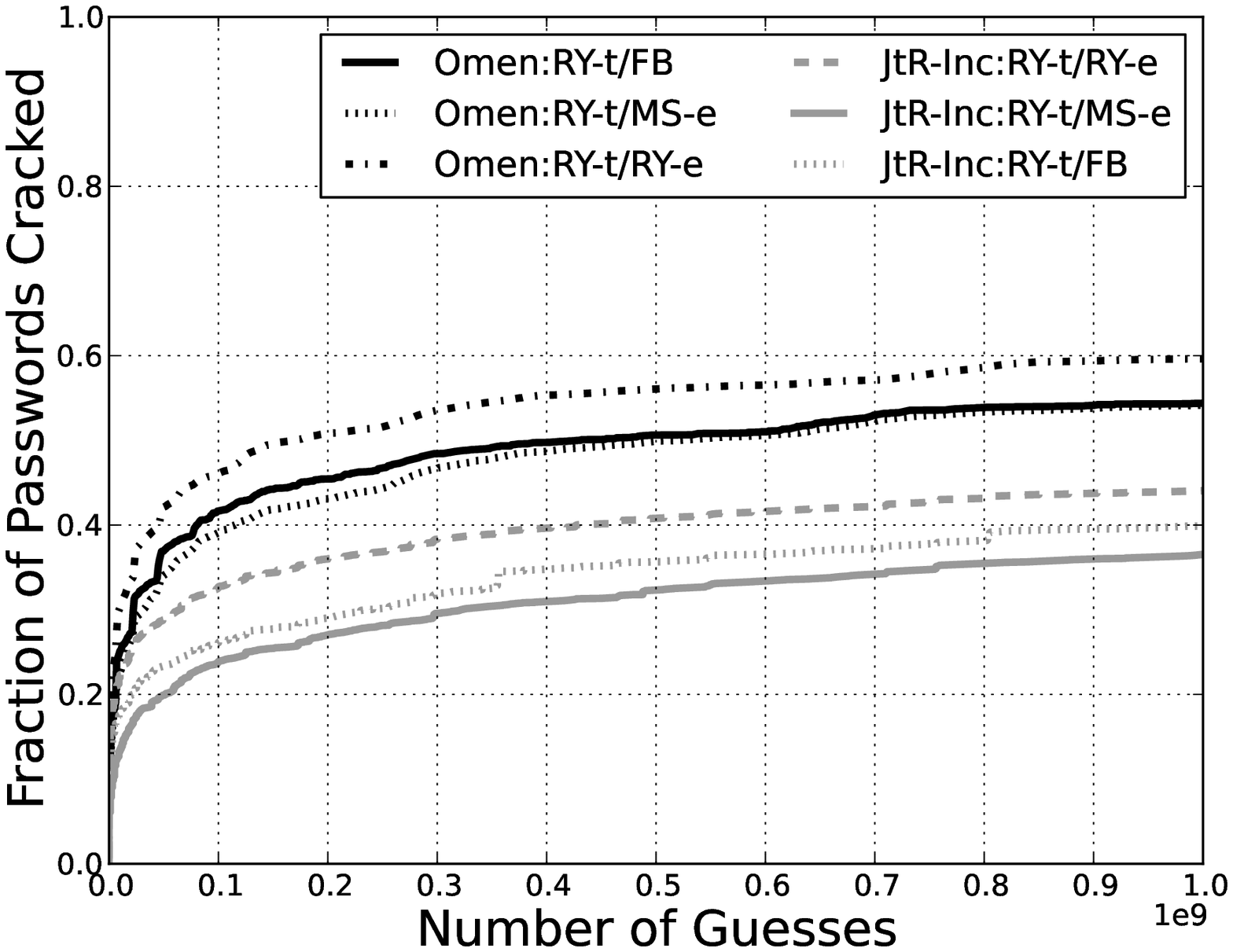}
  				\caption{}
  				\label{fig:ex-omen-vs-inc}
        \end{subfigure}
        \caption{Comparing OMEN to the PCFG guesser (a) and to JtR incremental mode(b).}
\vspace*{-1cm}
\end{figure}


\paragraph{OMEN vs PCFG}
Figure~\ref{fig:ex-omen-vs-pcfg} compares our guesser with the PCFG
password guesser of Weir et al.~\cite{weir-2009-pwd-crack-grammars},
based on the code available in~\cite{weir-website}.
We run it using the configuration as described in the paper and using
the dictionary dict-0294~\cite{dict-0294}.
In this experiment we trained on the MySpace dataset, since the public version
of PCFG seemed unable to correctly train on the larger RockYou dataset due to a memory bug. 

OMEN outperforms the PCFG guesser,
except when testing on the MySpace list, where PCFG produces slightly better guesses
for the first 100 million attempts. However, OMEN produces better guesses after 100 million and outperforms PCFG by around 10\% at
8~billion guesses. We believe the reason is that the grammar for
PCFG is trained on a subset of the MySpace list, which is
better adapted to guessing the MySpace list, whereas the MySpace list
is too small to meaningfully train Markov models.  %
Note that PCFG mostly plateaus after $0.5$ billion guesses
and results hardly improve any more, whereas OMEN still produces
noticeable progress.
For the other testing sets, however, OMEN produces better guesses almost from the beginning.

%
%
%

\paragraph{OMEN vs JtR's Incremental Mode}
We also compare OMEN to JtR in incremental mode (Figure \ref{fig:ex-omen-vs-inc}).
Similarly to the previous experiments, both crackers were trained on the RockYou training set
of $30$ million passwords. It appears clear that the incremental mode in JtR 
produces worse guesses than OMEN. Notably, it also produces worse guesses than any other
cracker tested.




\section{Personal Information and Password Guessing}
\label{sec:personal-info}

The results of the previous section show that a significant fraction of passwords can be guessed with a (relatively) moderate
number of attempts, compared to the entire possible search space. However, most techniques
adopt a {\em coarse grained}  approach that relies on a {\em generic} probability distribution, which 
by definition, does not depend on the password being guessed.  Intuitively, exploiting personal information in the password cracking process may enhance the success ratio.  Surpassingly, such possibility has not been extensively studied.

%
While a multitude of personal information can be used, we focus on a realistic scenario where these information can easily be
extracted from a \textit{public} source. For instance, an attacker armed with his victim email address, can gather her social network profile and use the collected information to guess her password. Such information includes:
\begin{itemize}
\item Information related to the \emph{user's name}, such as first
  name, last name, username;
\item \emph{Social relations} such as friends' and family members' names;
\item \emph{Interests} such as hobbies, favorite movies, etc;
\item \emph{Location information} like the place of residence;
\end{itemize}

In the next sections, we explore the relationship between such
information, referred to as {\em hint} in the rest of this paper,  and passwords, as well as 
its effect on password cracking.

\subsection{Similarity between Passwords and Personal Information}
\label{sim_sec}
To asses whether social information can be exploited to improve password cracking, we quantify the correlation between passwords and personal information. We use two different similarity metrics to capture different aspects of the potential overlap. 

\descr{Longest common substring (LCSS):}
The LCSS of two strings is the
longest string that is a substring of both strings.
For example, the LCSS of the two strings \texttt{abcabc} and
\texttt{abcba} is \texttt{abc}, and the LCSS of \texttt{abcabc}.
%



\begin{wrapfigure}{r}{55mm}
  \centering
  \includegraphics[scale=0.3]{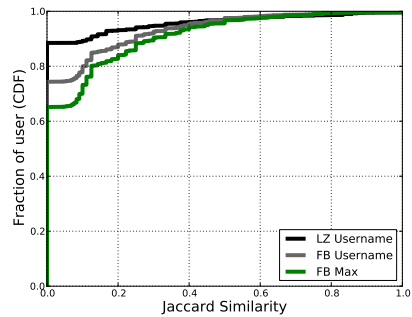}
   \caption{CDF of Jaccard similarity}
   \label{fig:cdf_sim}
\end{wrapfigure}

\descr{(Modified) Jaccard similarity (JS):}
The Jaccard similarity 
index compares similarity of two sets.  For two sets $X$ and $Y$, the
Jaccard index is $J(X,Y) := \frac{|X \cap Y|}{|X \cup Y|}$.
It is a similarity measure on sets, not on strings.  However,
 we extract the $n$-grams from a string and apply the
JS function to the sets of $n$-grams.
There is one drawback of this measure that does not match our
application, namely that appending unrelated information to the hint
(which degrades the ``real'' usefulness only for large appended text)
rapidly decreases the JS value.
Therefore, we use a ``modified JS'' defined as follows:
Given a password $P$ and a \emph{hint} $H$, and denoting the set of
$3$-grams that appear in $P$ and $H$ with $P_{3g}$ and $H_{3g}$, respectively, we define
$J^*_{3g}(P,H) := \frac{ | P_{3g} \cap H_{3g} | }{ |P_{3g}| }.$
Figure \ref{fig:cdf_sim} displays the  cumulative distribution function (CDF) of the JS between passwords and personal information 
for FB and LZ datasets. For each password of the Facebook dataset, we compute the JS with each of its corresponding personal attribute.

The lowest (green) plot (FB Max) displays the CDF of the maximum of these values. It basically shows that about 35\% of Facebook 
passwords are somewhat correlated with one
of their user's attributes. Note that any non-zero similarity means that at least one $3$-gram is shared, which already is a substantial overlap.
The correlation becomes stronger for about 10\% of users. The grey  (FB Username) and black (LZ Username) curves display the CDF of the similarity 
between passwords and $UserNames$ for the FB and LZ datasets, respectively. They both show similar shape, although surprisingly Facebook passwords seem to 
be more correlated to $UserNames$ than LZ passwords. In both datasets, about 10\% of passwords seem to be highly correlated with the $UserNames$ attribute.

\begin{table}[ht!]
\begin{center}
\begin{tabular}{|c||c|c|c|c|c|}
\hline Attribute 		& JS	& JS(5\%) 	& LCSS  &  LCSS(5\%)	& Len\\
\hline FirstName		&0.02	&0.31   &0.93		&4.34 	& 	5.84\\
\hline LastName			&0.01	&0.24   & 0.71	&3.55 	&	6\\
\hline Username			&0.07	&0.58   &1.48		&6.31 &	10.53 \\
\hline Friends			&0.06	&0.30   &1.54		&4.15 &	147.28\\
\hline Edu/Work		&0.02	&0.23   & 1.20	&3.5 &		40.93\\
\hline Contacts			&0.06	&0.63    &1.44	&6.55 &	17.67\\
\hline Location			&0.01	&0.13    &1.07	&2.94 &	25.70\\
\hline Birthday			&0.04	&0.5     & 0.87	&4 &		6.84\\
\hline Siblings			&0.04	&0.36     & 1.27	&4.96 	&	94.71\\
\hline 
\end{tabular} 
\caption{Mean similarity between passwords and personal information (FB dataset).}
\label{sim_tab}
\end{center}
\vspace*{-1cm}
\end{table}

Table \ref{sim_tab} goes into more details and summarizes similarity measures between different attributes and the respective passwords of the FB datasets.
As shown by Figure \ref{fig:cdf_sim}, more than 80\% of passwords have little similarity with personal information attributes. This explains the small similarity values 
in column JS and LCSS, that correspond to averages of the similarities over the whole dataset (since many similarities are equal to zero, the resulting averages have
pretty low values).
For this reason, we order the passwords according to their similarity values for the different attributes. We then present, in the columns JS(5\%) and LCSS(5\%), the average similarity of the top 5\% for each attributes.



First we notice how attributes such as $UserNames$, $FirstName$ and $Birthday$ seem to substantially overlap with the passwords.
For instance, for the top 5\% users, $UserNames$ and $Birthday$ share half of the $n$-grams with the password and have more
than 4 and 6.4 common substring with it respectively. 
Furthermore, we notice that long attributes such as $Friends$ or $Education$ and $Work$ (on average 150 characters long and 50, respectively) 
have a high value of LCSS. 
 Finally, the LCSS(5\%) results show that the average LCSS value is around 3 which sustains the usage of a 3-grams model (rather than 2-grams or 4-grams).


In Section \ref{sec:estimating-alpha} we will explore how to incorporate these findings to robustly increase the 
performance of OMEN.


\subsubsection{Password Creation Policies and Usernames}


One surprising fact highlighted by the data in the previous section is
that usernames and passwords are very similar in a small, yet significant,
fraction of the cases. This fact prompted us to study this specific aspect of password policies
in depth, reader may refer to Appendix~\ref{annex1} for more details. The results are worrisome: out of 48 tested sites, 27 allowed \emph{identical} username and password, including
major sites such as Google, Facebook, and Amazon, and only 4 sites
required more than one character difference between the two. This could lead to highly effective guessing attack.

\subsection{OMEN+: Improving OMEN Performance with Personal Information}
\label{sec:adding-personal}

In Section~\ref{sim_sec}  we showed that
users' personal information and passwords can be correlated.  However, it is not clear
how to  use such information when guessing passwords especially that some information may have a negatively 
impact on the performance.  Let us illustrate this possibility with an example: assume that we possess extensive information about a victim. This information may include name, date of birth, location information, family member names, etc.
Intuitively, this information should increase the performance of a password cracker. For example,
we could generate a password guess with the name of a sibling concatenated with their year of birth.
However, in order to increase the (overall) performance, one must still order password guesses 
in decreasing probability. Otherwise, the integration of additional information can decrease effectiveness.
To show this, for the sake of argument, let us assume that the attacker is only allowed 
{\em one} guess. The same argument can be extended to any number of guesses. 
The attacker should use some personal information for the one guess only if the probability of 
this password is higher than the most frequent ``generic'' password, say \texttt{123456}.
Assuming that no user specific password is, on average, more frequent than \texttt{123456},
then, by including personal information, the attacker would decrease her chances of success
for the single guess.

\paragraph{Boosting Algorithm}
%
%
It is challenging to decide how to use the additional personal
information. In fact, only certain parts of this data
overlap with the password.  In a dictionary-based attack, we need
to decide which substring(s) should be added to the attack dictionary,
and choosing the wrong ones one could decrease performance.
With Markov models, however, the situation is easier, as $n$-grams
are a canonical target.  By increasing (conditional) probabilities
of important $n$-grams (i.e. $n$-grams that are contained in $hints$), 
 we can increase the probability of  passwords that are related to them, and thus improve
 OMEN's performance.
Our boosting algorithm takes as input a parameter $\alpha > 1$ (see
Section~\ref{sec:estimating-alpha} on how this parameter is chosen), a
hint $h$, and a Markov model consisting of the initial probabilities
$p(xy)$ and the conditional probabilities $p(z|xy)$, and outputs
modified conditional probabilities $p^*(z|xy)$.
Let us assume we have a list of $N$ passwords $pwd_1, \ldots, pwd_N$, and for
each password $pwd_i$ we have some additional information $hint_i$,
which may or may not help us in guessing the password.
We want to automatically and efficiently determine if a specific set
of hints is useful or not, and how strongly each of the hints should be weighted. 
Note that since hints are password-specific, trying all possible combinations
would be too computationally expensive.
%
%
%
Our algorithm works as follows\footnote{To ease presentation, we only describe the estimation algorithm for $3$-grams. 
The generalization to $n$-grams is straightforward.}:
\begin{enumerate}
\item For each pair $pwd_i, hint_i$, two sets are defined:
$S_i$ is the set of $3$-grams that appear in both the password and the
hint
 and $T_i$ is the set of $3$-grams from $pwd_i$ such that $hint_i$ has
a $3$-gram that shares the first two letters, but not the third.  For
instance, if $pwd_i = \mathtt{password}$ and $hint_i =
\mathtt{passabcd}$ then $S_i = \{ \mathtt{pas}, \mathtt{ass} \}$
and $T_i = \{\mathtt{ssw}\}$.

\item For each $3$-gram
$xyz$ in $S_i$, we set the conditional probability
\[
p^*(z|xy) := \alpha \cdot p(z|xy)
\]
for a given parameter $\alpha$.  
Considering the previous example, we boost the $3$-grams $\mathtt{pas}$ and $\mathtt{ass}$,
as follows:
$
p^*(s|pa) := \alpha \cdot p(s|pa)  ;
p^*(s|as) := \alpha \cdot p(s|as)
$

By modifying a conditional probability from $\hat p$ to $\alpha \cdot
\hat p$ 
we distribute a probability mass of $(\alpha - 1)
\hat p$ that we need to subtract at another place.  The
probability $\hat p$ is (in practice) much smaller than $1$ (we use an
alphabet size of $|\Sigma| = 72$), so $1 - \hat p \approx 1$.  Consequently,  
if we multiply all remaining (conditional) probabilities except
$\hat p$ with $(1 - \alpha \hat p)$, they sum up to approximately $1$ again (using the approximation simplifies the calculations):
$(1 - \alpha \hat p) \cdot ( \sum_{i \neq z} p(i|xy) ) + \alpha \hat p \approx (1 - \hat p \alpha) \cdot 1 + \alpha \hat p = 1.$
%
Writing $s_i := |S_i|$ and $t_i := |T_i|$ for the sizes of the two
sets, the overall effect on password probabilities is
\[
p^*_{pwd_i} = \prod_{i \in (pwd_i)_{3g}} p_i \approx \alpha^{s_i} (1 - \hat  p \alpha)^{t_i} \cdot p^{old}_{pwd_i}
\]
where $p^*_{pwd_i}$ is the ``new'' probability after boosting the
$n$-grams, and $p^{old}_{pwd_i}$ is the ``old'' probability before
boosting.

%

\end{enumerate}

\subsubsection{Estimating Boosting Parameters}
\label{sec:estimating-alpha}

\begin{figure*}[ht!]
  \centering
  \begin{minipage}{0.45\linewidth}
    \includegraphics[width=\linewidth]{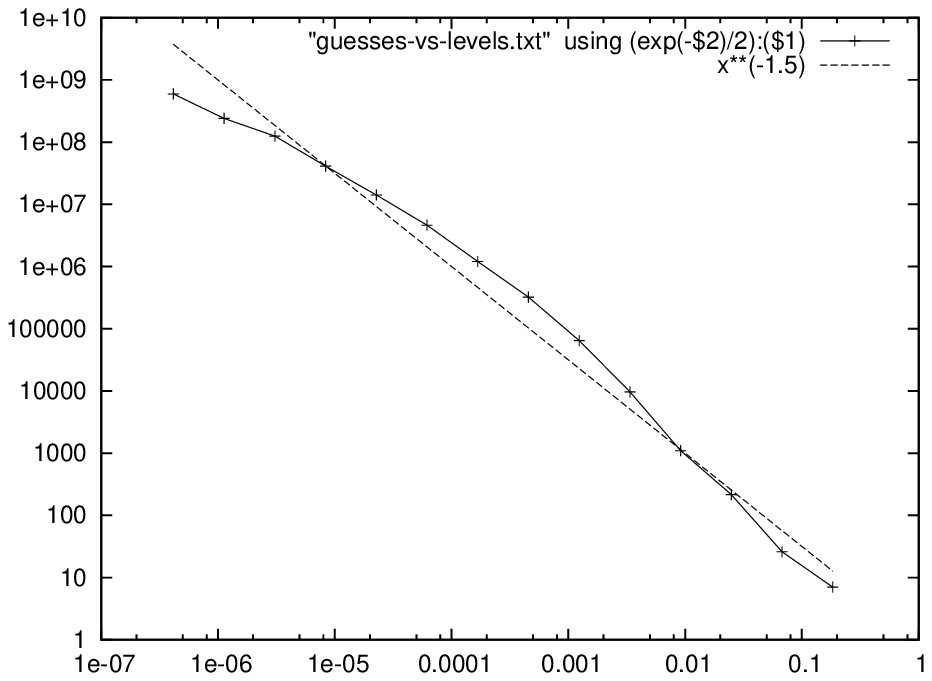}
     \caption{Relation between (estimated)
      password probability and the position when it is guessed (line), and the
      approximation we use (dashed-line), on double-logarithmic scale.}
    \label{fig:guess-vs-lvl}
  \end{minipage}
  \hspace*{0.03\linewidth}
  \begin{minipage}{0.45\linewidth}
   \centering
   \begin{tabular}{l | ll | l}
                     &$\alpha$        &$\ln(\alpha)$    & boosted\\
     \hline
     email           &1.6             &0.5               &0(*)\\
     userName        &2               &0.7               &1\\
     firstName       &2.3             &0.8               &1\\
     lastName        &1.5             &0.4               &0\\
     birthday        &5               &1.6               &2\\
     location        &1.7             &0.5               &1\\
     contact         &1.5             &0.4               &0\\
     eduWork         &1.1             &0.1               &0\\
     friends         &1.4             &0.3               &0\\
     siblings        &1.7             &0.5               &1\\
   \end{tabular}
    \captionlistentry[table]{la}
    \captionsetup{labelformat=andtable}
    \caption{The estimated values of $\alpha$ and the boosting parameters for the attributes we considered.}
   \label{tab:estimated-alphas}
  \end{minipage}
 
\end{figure*}

The section describes how the boosting parameter $\alpha$ of each $hint_i$ is computed.
Recall that OMEN outputs password guesses in descending order of their
(estimated) probabilities.  Let $f$ denote the function that gives the
estimated probabilities $x=f(y)$ for the $y$-th guess that OMEN
outputs.
This function can simply be computed by running OMEN and printing
the probability estimation of the current password. 
The inverse function $y = f^{-1}(x)$ gives the number of guesses
OMEN needs to output before reaching passwords with a certain
(estimated) probability.
This function is shown in Figure~\ref{fig:guess-vs-lvl} on a double
logarithmic scale.  In order to simplify the subsequent calculations,
we approximate this function as $f^{-1}(x) \approx x^{-1.5}$.

%
%

The estimated number of guesses which is  required to crack \emph{all} passwords
$pwd_1, \ldots, pwd_N$ is consequently defined by
$S=\frac{1}{N} \sum_{i=1,\ldots,N} f^{-1}( p_{pwd_i} ).$%
%
Therefore, for a given $hint_i$, the value $\alpha_i$ to use to boost this hint is the value of $\alpha$ that minimizes the following function:
$
S^* = \frac{1}{N} \sum_{i=1,\ldots,N} ( p^*_{pwd_i} )^{-1.5}.
$

The following section presents the optimal values of $\alpha$ for different hints.
\begin{figure}
        \centering
        \begin{subfigure}[b]{0.3\textwidth}
                \centering
                \includegraphics[width=\textwidth]{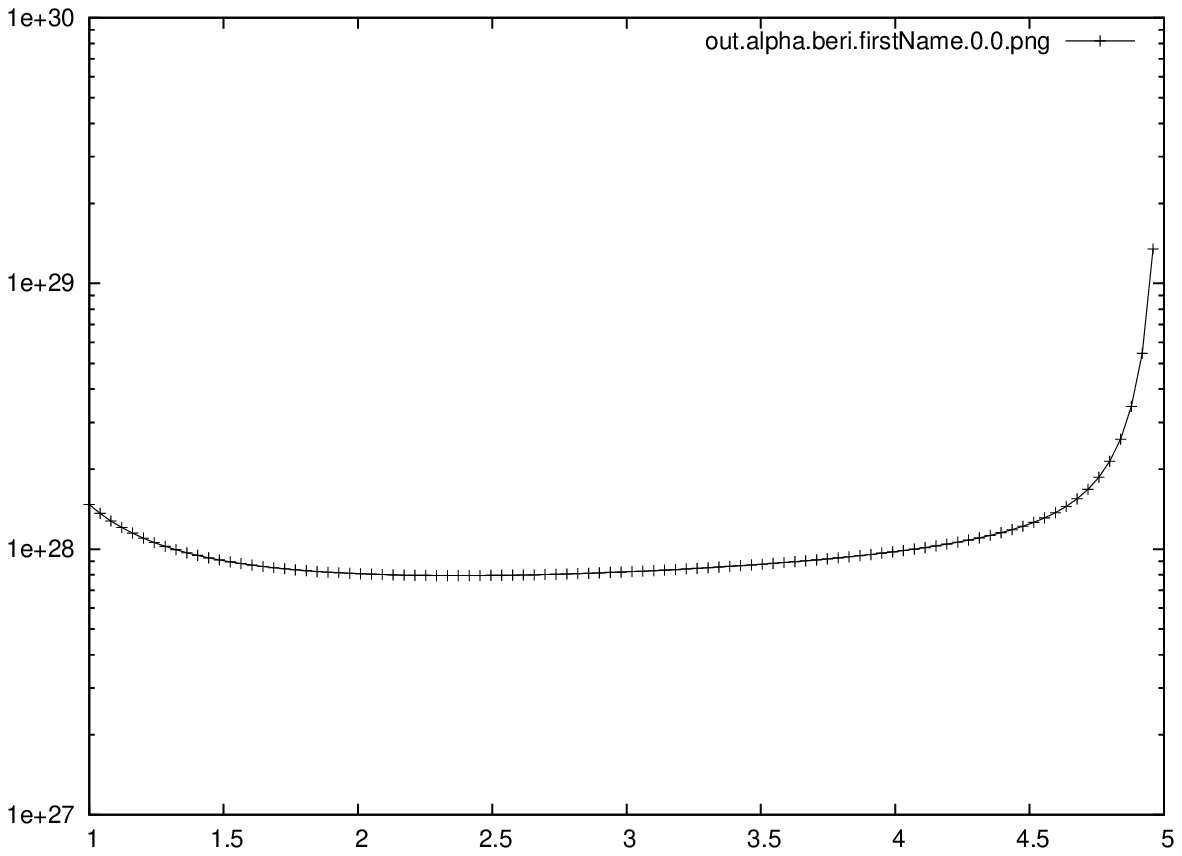}
                \caption{Username}
                \label{fig:alpha-measured-firstName}
        \end{subfigure}%
        ~ 
        \begin{subfigure}[b]{0.3\textwidth}
                \centering
                \includegraphics[width=\textwidth]{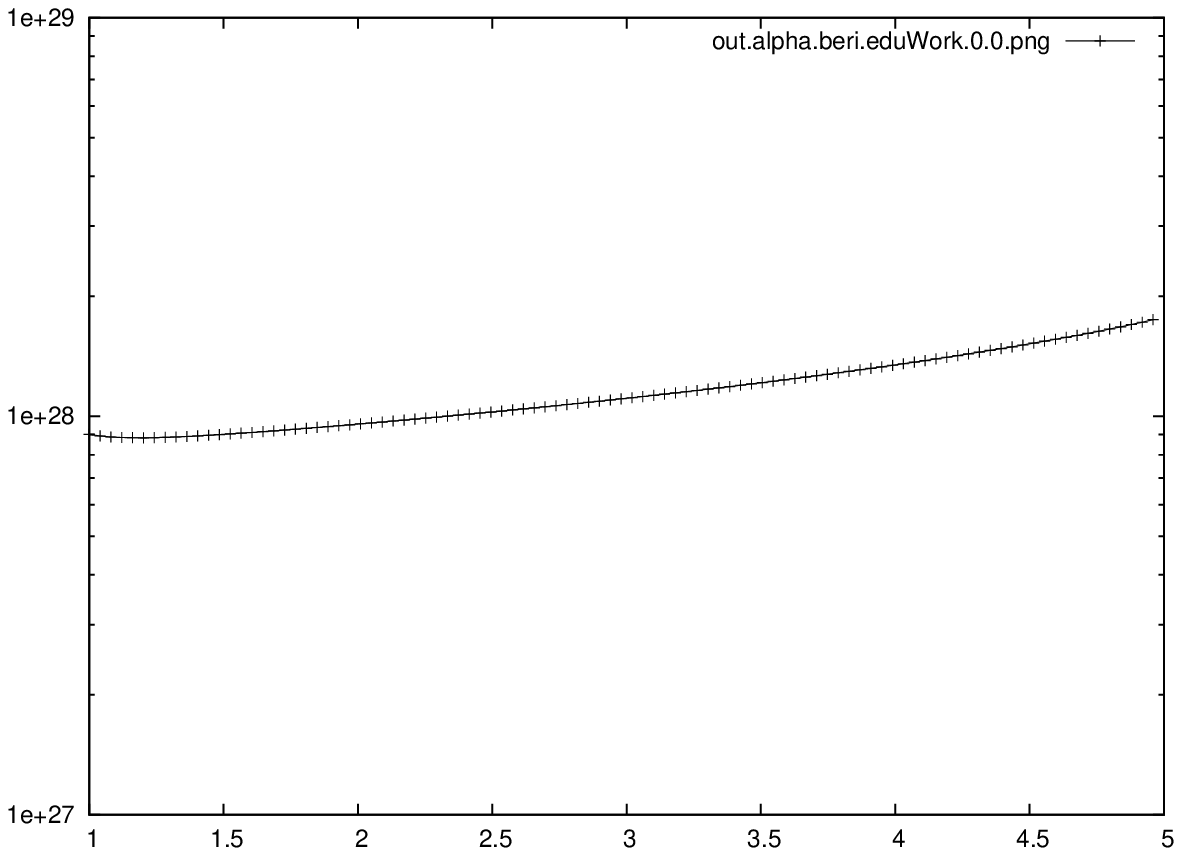}
                \caption{EduWork}
                \label{fig:alpha-measured-eduWork}
        \end{subfigure}
        ~ 
        \begin{subfigure}[b]{0.3\textwidth}
                \centering
                \includegraphics[width=\textwidth]{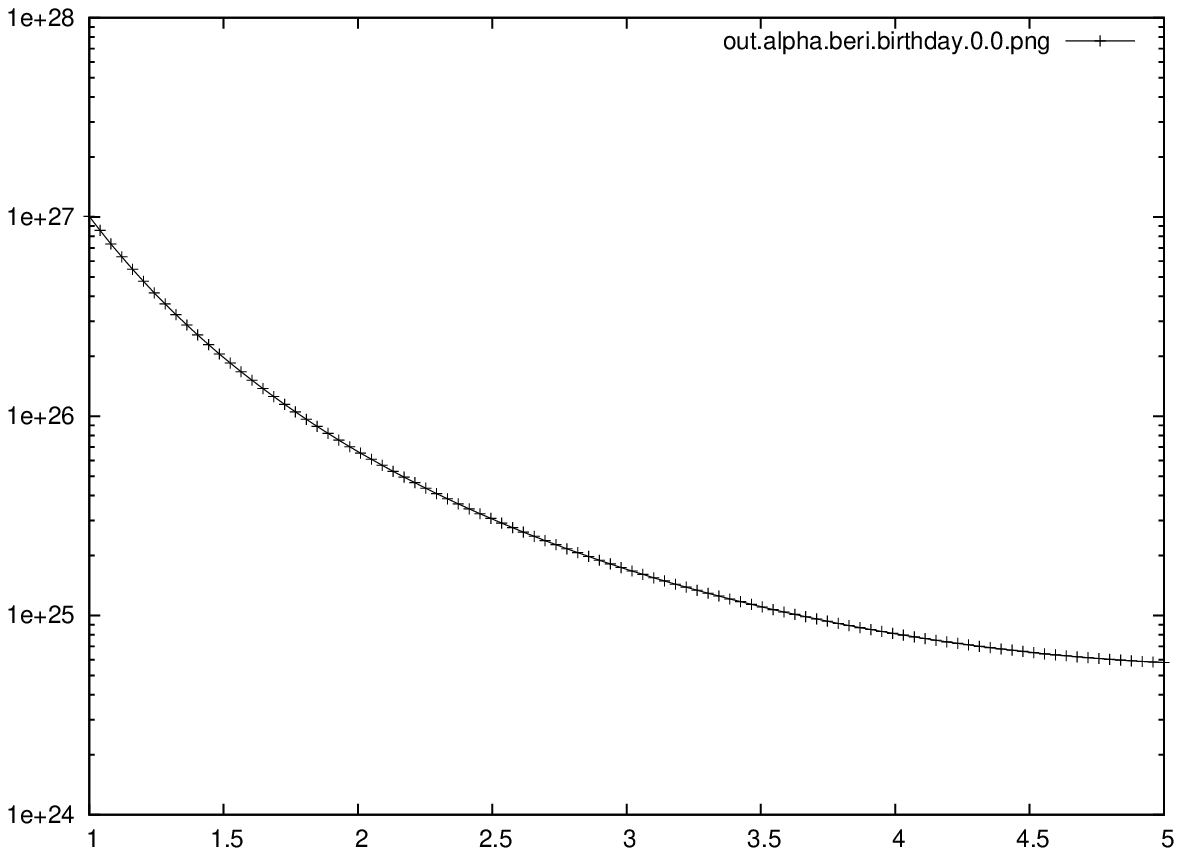}
                \caption{Birthday}
                \label{fig:alpha-measured-birthday}
        \end{subfigure}
        \caption{Estimated influence of the parameter $\alpha$ applied on the
      attribute: \emph{username} (a), \emph{eduWork} (b) and \emph{birthday} (c).  The $x$-axis shows $\alpha$ and the
      $y$ axis the expected number of guesses.}
        \label{fig:boostingFig}
\vspace*{-0.6cm}
\end{figure}
%

%

\subsection{Evaluation}
\subsubsection{Boosting Parameter Estimation}
We use the techniques described in the previous section to estimate
the boosting parameter $\alpha$ for the different hints. For each hint,
we compute the sum $S^*$  for different values of $\alpha$, and select the value that minimizes it.
We illustrate the results with three examples:\\
\descr{First Name: }Figure~\ref{fig:alpha-measured-firstName} shows $S^*$ as a function of $\alpha$ for the
attribute \emph{first name}. The minimum is around $\alpha = 2.3$, which yields a boosting parameter of $1$.

\descr{EduWork: }Figure~\ref{fig:alpha-measured-eduWork} shows  $S^*$ as a function of $\alpha$ for the
attribute \emph{eduWork}, which is an identifier that contains the persons'
education and occupancy.  It has a very small decrease in the
beginning, with a minimum around $\alpha=1.2$, but the overall
differences are small and the remainder of the graph is monotonically
increasing.  The boosting parameter is $0.1$, which is rounded to $0$.

\descr{Birthday: }Figure~\ref{fig:alpha-measured-birthday} shows $S^*$ as a function of $\alpha$ for the
\emph{birthday} attribute.  Note that our dataset only contains
a small number of profiles with that attribute, so the result might  not be
necessarily meaningful.  Overall, there were only 7 profiles where the
birthday attribute have an effect, and for two of them the effect is
positive.  These two have enough effect to lead to a great advantage
in using the attribute.  Overall, we limited the maximal parameter
considered to 5.

Finally, we dropped the attribute \emph{email}. Although it
gives an alpha of $1.6$ since the \emph{username} is
contained in email and achieves better results. All boosting parameters are summarized in Table~\ref{tab:estimated-alphas}. 

\begin{figure}[h!]
        \centering
        \begin{subfigure}{0.5\textwidth}
               \centering
  			   \includegraphics[width=\linewidth]{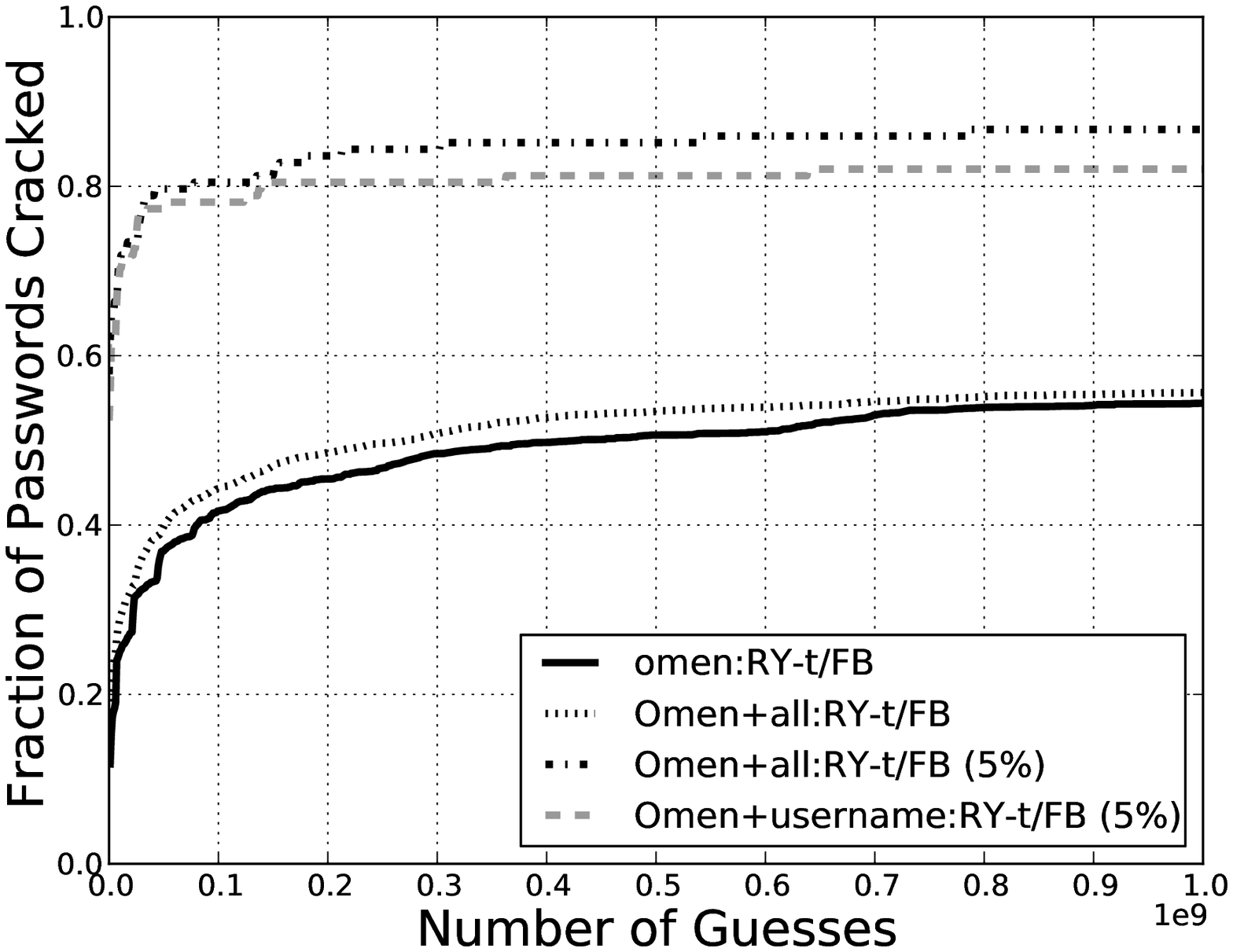}
 			   \caption{Comparing OMEN with and without personal information on the FB list }
 				\label{fig:nemo-fb}
        \end{subfigure}%
        ~ 
        \begin{subfigure}{0.5\textwidth}
            \centering
            \includegraphics[width=\linewidth]{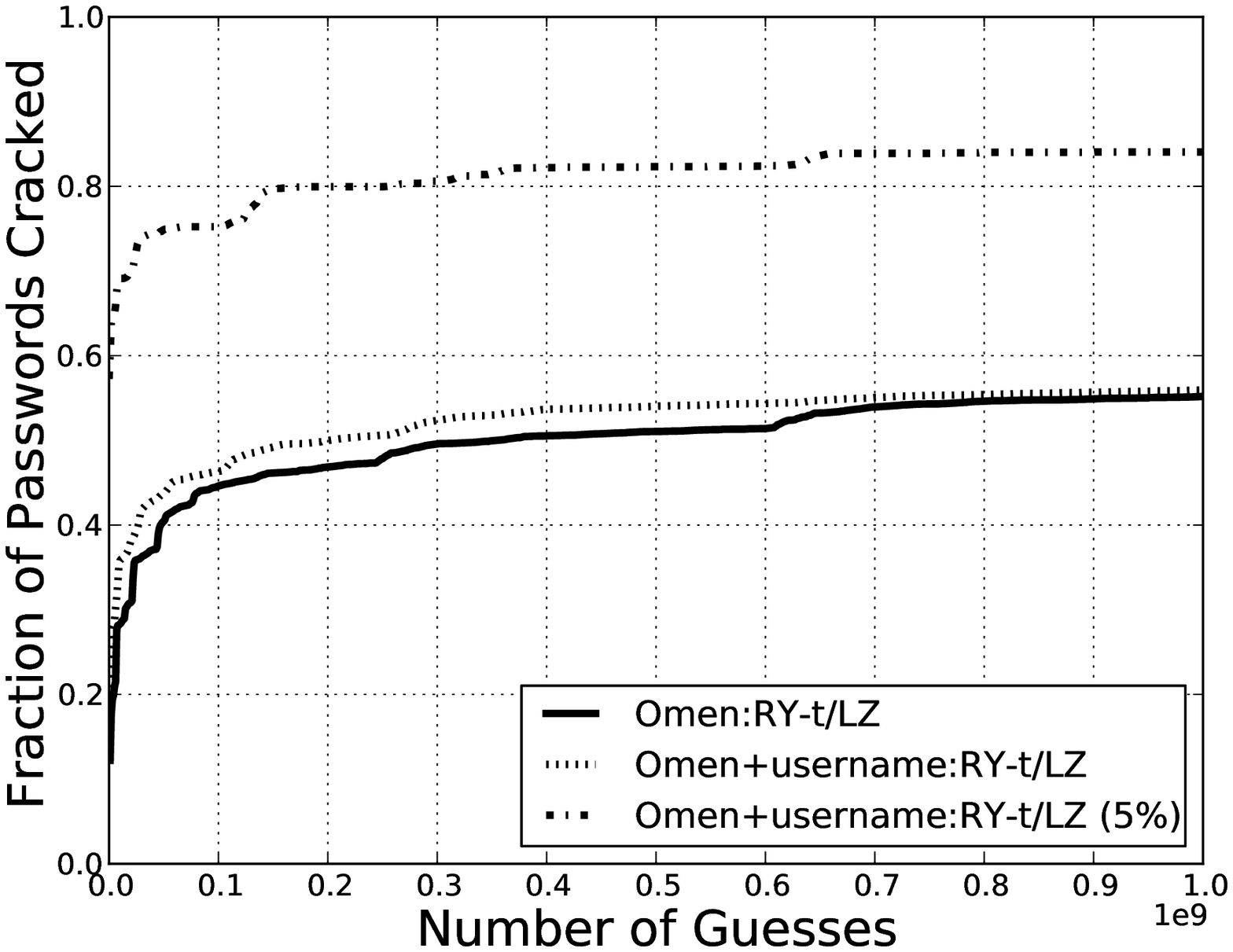}
            \caption{Comparing OMEN with usernames as hint and without on the LZ/FB list }
            \label{fig:nemo-lz}
        \end{subfigure}
\vspace*{-0.6cm}
\end{figure}

\vspace*{-0.5cm}
\subsubsection{OMEN+ Performance}
Once we have estimated the values of the parameters $\alpha_i$ for each $hint_i$, we 
run OMEN+ on the Facebook (FB) list, consisting of $3140$ passwords together
with publicly available information about the users. The results are 
presented in Figure \ref{fig:nemo-fb}. As expected, by including personal
information in the Markov model, OMEN+ is able to guess more passwords in absolute terms.
We have also conducted different experiments with other values $\alpha_i$ to test
the effectiveness of our estimation code. We confirmed that, when using different values $\alpha_i$,
the cracking performance either remains the same or slightly decreases. The two lower curves in 
the figure show the performance of OMEN+ over all passwords with and without using personal attributes. Using personal information
can increase the guessed passwords up to $5\%$ (for lower number of guesses of up to 100 million), and around $3\%$ at 1 billion guesses.
The limited performance gain is partially explained by the fact that, as shown in Section \ref{sim_sec}, only a small proportion of passwords are based on personal information.
The 2 curves in the upper part of the figure display  the performance of OMEN+ over the top 5\% passwords that are the most
correlated with their personal attributes. The achieved performance is much better: About 82\% of the passwords are cracked by using
usernames only, and more than 88\% by using all considered personal attributes. This result is very promising.Figure \ref{fig:nemo-lz} shows a similar experiment performed on the LZ list (60000 passwords).
We only had access to the local-part (i.e., the username) of the email associated to each password.
Even though the information in this case was more limited, we realized similar gains compared
to the previous test on the more extensive FB data.


%

%
%
%

\vspace*{-0.4cm	}
\section{Discussion and Conclusion}
\label{sec:discussion}
\vspace*{-0.3cm	}
In this work we have first presented an efficient password guesser
(OMEN) based on Markov models, which outperforms all publicly available password guessers.  For common
password lists we found that we can guess almost 70\% of the passwords with 10 billion
guesses.
%
Subsequently, in our second contribution, we tested if additional
personal information about a user can help us to better guess
passwords.  We found that some attributes indeed help, and we showed
how OMEN+ can efficiently leverage this information.  
%
We summarize some of the key insights:
\begin{itemize}
\item Our work shows that Markov Model have even
  better potential than previously thought~\cite{ccs05}, as we could make them
  guessing ``in order'', which leads to the improvements shown in
  Figures~\ref{fig:ex-omen-vs-jtr-markov}
  and~\ref{fig:10B-omen-vs-jtr-markov}.


\item We find, with our preliminary and simple experiments, that we can guess up to 5\% more passwords. However this percentage is a lower bound
since we had access to only a limited number of personal information and attributes. Furthermore, we show that the gain can go up to 30\% for 
  passwords that are actually based on personal attributes. This result clearly shows that passwords based on personal information are weaker and
  should be avoided. 



\end{itemize}

\newpage

\bibliographystyle{acm}
\bibliography{bibfile,pwd-hw}

\newpage
\appendix
\makeatletter
\def\@seccntformat#1{Appendix~\csname the#1\endcsname:\quad}
\makeatother
\section{Password creation policies}
\label{annex1}
One surprising fact highlighted by the data in the previous section is
that usernames and passwords are very similar in a small, yet significant,
fraction of the cases.
Common sense mandates that a password should not be too similar to the
corresponding username, because the username is 
 almost always available to the attacker in a guessing
attack.  

This fact prompted us to study this specific aspect of password policies
in more detail.
We conducted a
brief test\footnote{This survey is neither representative nor
  complete, but the results are clear enough to show that the problem
  exists on a large scale.} across $48$ popular international sites (from the Alexa Top 500 list),
to see how they handle similarities between the username and the
password (See Table~\ref{tab:pwd-survey}).
%

These sites have different demands for security, ranging from
relatively low security demands (Facebook, Twitter), to high security
demands (Ebay, PayPal).  We did not rely on the stated password
policies, but manually created an account on each site.
We created a random but plausible username that was not yet used with
the service, and tried to register an account.  We initially tried to
use the username as the password, and if that failed we tried
subsequent modifications until we succeeded.
The results are worrisome, but not too surprising.  Out of 48 sites
tested, 27 allowed \emph{identical} usernames and passwords, including
major sites such as Google, Facebook, and Amazon, and only 4 sites
required more than one character difference between the two.
This could lead to highly effective guessing, for example in the
Lulzsec-dataset, we found that 5\% of the accounts had strongly overlapping usernames
and passwords.  



\thispagestyle{empty}
\pagestyle{empty}

\begin{table*}[p]
  \centering
  \footnotesize
  \begin{tabular}{l | ll | ll | l}
    Account             &Username               &Password	&\parbox[b]{1.5cm}{Same\\accepted?}	
                                                                        &\parbox[b]{1.5cm}{Min.\ Diff.\ (\# Chars)}	
                                                                                                &Comments\\
    \hline
    Google              &berkusrnfe02@gmail.com	&berkusrnfe02	&yes	&0	\\	
    Facebook            &berkusrnfe02@gmail.com	&berkusrnfe02	&yes	&0	\\	
    Twitter             &berkusrnfe02           &berkusrnfe03	&no	&1	\\
    Baidu               &berkusrnfe02           &berkusrnfe03	&no	&1	\\	
    Ebay                &berkusrnfe03           &BerkUsrnfe14	&no	&2	&\parbox[t]{4cm}{Username cannot be same as email; requires capitals.}\\
    Amazon              &berkusrnfe02@gmail.com	&berkusrnfe02	&yes	&0	\\	
    Paypal              &berkusrnfe02@gmail.com	&berkusrnfe03	&no	&1	\\	
    Yahoo               &berkusrnfe02@yahoo.com	&berkusrnfe03	&no	&1	\\	
    Wikipedia           &berkusrnfe02           &berkusrnfe03	&no	&1	\\	
    Windows Live        &berkusrnfe02@hotmail.com &berkusrnfe03	&no	&1	\\	
    \hline
    QQ.com              &berkusrnfe02           &berkusrnfe02	&yes	&0	\\	
    LinkedIn            &berkusrnfe02@gmail.com	&berkusrnfe02	&yes	&0	\\	
    Taobao              &berkusrnfe02           &berkusrnfe02	&yes	&0	\\	
    Sina.cn.com              &berkusrnfe02@yahoo.com	&berkusrnfe02	&yes	&0	\\	
    MSN                 &berkusrnfe02@gmail.com	&berkusrnfe03	&no	&1	\\	
    WordPress           &berkusrnfe02           &berkusrnfe02	&yes	&0	\\	
    Yandex              &berkusrnfe02           &berkusrnfe03	&no	&1	\\	
    163.com             &berkusrnfe02@163.com	&berkusrnfe03	&no	&1	\\
    Mail.ru             &berkusrnfe02@Mail.ru	&berkusrnfe03	&no	&1	\\
    Weibo               &berkusrnfe02@gmail.com	&berkusrenfe02	&no	&0	\\
    \hline
    Tumblr              &berkusrnfe02           &berkusrnfe02	&yes	&0	\\	
    Apple               &berkusrnfe02           &BerkUsrnfe02	&no	&0	&\parbox[t]{4cm}{Password at least 1 capital, 1 number, no 3 consecutive identical characters, not same as account, at least 8 char.}\\
    IMDB                &berkusrnfe02           &berkusrnfe02	&yes	&0	\\	
    Craigslist          &berkusrnfe02@gmail.com	&berkusrnfe03	&no	&1	\\	
    Sohu                &berkusrnfe02@gmail.com	&berkusrnfe02	&yes	&0	\\	
    FC2                 &berkusrnfe02           &berksurnfe03	&no	&3      &\parbox[t]{4cm}{The password cannot contain any five (5) consecutive characters of your e-mail address.}\\
    Tudou               &berkusrnfe02@gmail.com	&berkusrnfe02	&yes	&0	\\	
    Ask                 &berkusrnfe02           &berkusrnfe02	&yes	&0      &\\
    iFeng               &berkusrnfe02           &berkusrnfe03	&no	&1	\\	
    Youku               &berkusrnfe02           &berkusrnfe02	&yes	&0	\\	
    \hline
    Tmall               &berkusrnfe02           &berksurnfe03	&no	&3	\\	
    Imgur               &berkusrnfe02           &berkusrnfe02	&yes	&0	\\	
    Mediafire           &berkusrnfe02@gmail.com	&berkusrnfe02	&yes	&0	\\	
    CNN                 &berkusrnfe02           &berkusrnfe02	&yes	&0	\\	
    Adobe               &berkusrnfe02           &berkusrnfe02	&yes	&0	\\	
    Conduit             &berkusrnfe02@gmail.com	&berkusrnfe02	&yes	&0	\\	
    odnoklassniki.ru/   &berksrnfe02            &berkusrnfe03	&no	&1	\\	
    AOL                 &berkusrnfe02           &beruksrnef03	&no	&5	\\	
    The Pirate Bay      &berkusrn               &berkusrn	&yes	&0      &\parbox[t]{4cm}{Username length limit.}\\
    ESPN                &berkusrnfe02           &berkusrnfe02	&yes	&0	\\	
    \hline
    Alibaba             &berkusrnfe02           &berkusrnfe02	&yes	&0	\\	
    Dailymotion         &berkusrnfe02           &berkusrnfe02	&yes	&0	\\	
    Chinaz              &berkusrnfe02           &berkusrnfe02	&yes	&0	\\	
    AVG                 &berkusrnfe02@gmail.com	&berkusrnfe02	&yes	&0	\\	
    Ameblo              &berkusrnfe02           &berkusrnfe03	&no	&1	\\	
    GoDaddy             &berkusrnfe02           &berkusrnfe02	&yes	&0	\\	
    StackOverflow       &berkusrnfe02           &BerkUsrnfe03	&no	&1      &\parbox[t]{4cm}{Needs capitals or special characters.}\\
    4shared             &berkusrnfe02@gmail.com	&berkusrnfe02	&yes	&0	\\	
  \end{tabular}
  \caption{Detailed results from a small survey on $48$ large sites concerning their password policies.}
  \label{tab:pwd-survey}
\end{table*}



\end{document}